\documentclass[12pt]{iopart}

\usepackage{iopams}

\usepackage{graphicx}

\usepackage{graphicx}
\usepackage{epstopdf, epsfig}
\usepackage[table,xcdraw]{xcolor}

\usepackage{wrapfig}
\usepackage{lipsum}

\usepackage[utf8]{inputenc}
 
\renewcommand{\arraystretch}{1.5}

\usepackage{dblfloatfix}

\usepackage{dcolumn}% Align table columns on decimal point
\usepackage{bm}% bold math
\usepackage{amsfonts}

\usepackage{color}

\usepackage{array}

\usepackage{setspace}

\usepackage{graphicx}
\usepackage{ragged2e}

\usepackage[font=small,labelfont=bf]{caption}

\usepackage{enumitem}

\usepackage{multirow}

\usepackage[superscript,biblabel]{cite}
\usepackage{textcomp}

\usepackage{tabularx}
\usepackage{adjustbox}
\usepackage{booktabs}

\usepackage{multirow}
\usepackage[table,xcdraw]{xcolor}

\usepackage[utf8]{inputenc}

\usepackage[font=scriptsize,labelfont=bf]{caption}

\usepackage{enumitem}

\definecolor{bl}{rgb}{0.5, 0, 0}
\definecolor{cobalt}{rgb}{0.0, 0.28, 0.67}
\definecolor{darkcerulean}{rgb}{0.03, 0.27, 0.49}
\definecolor{egyptianblue}{rgb}{0.06, 0.2, 0.65}

\usepackage{hyperref}
\usepackage{xcolor}
\hypersetup{
    colorlinks = true,
    linkcolor = blue,
    citecolor = blue,
    urlcolor = blue
}

\usepackage{lineno}
\usepackage[superscript,biblabel]{cite}

\usepackage{graphicx}

\usepackage[T1]{fontenc}
\usepackage[utf8]{inputenc}

\usepackage{multirow}
\usepackage{tabularx}

\begin{document}

\title[Vortex phenomena in T-junctions]{On a model-based analysis of vortex formations and decay in flows through bio-inspired T-shaped cavities}

\author[Basu et al.]{\footnotesize{Sneham Das$^1$ and Saikat Basu$^{2, *}$}}
\vspace{0.5cm}

\address{$^1$~Department of Chemical Engineering, Jadavpur University, Kolkata, West Bengal 700032, India\\
        \vspace{2mm}
         $^2$~Department of Mechanical Engineering, South Dakota State University, Brookings, South Dakota 57007, United States\\
	     }
\ead{$^*$\,Saikat.Basu@sdstate.edu}

\date{}

%%%%%%%%%%%%%%%%%%%%%%%%%%%%%%%%%%%%%%%%%%%%%%%%%%%%%%%%%%%%%%%%%%%%%%%%%%%%%%%%%%%%%%%%%%%%%%%%%%%%%%%%%%%%%%%%%%%%%%%%%%%%%%%%%%%%%%%%%%%%%%%%%%%%%%%%%%%%

%\begin{document}

%\maketitle

%%%%%%%%%%%%%%%%%%%%%%%%%%%%%%%%%%%%%%%%%%%%%%%%%%%%%%%%%%%%%%%%%%%%%%%%%%%%%%%%%%%%%%%%%%%%%%%%%%%%%%%%%%%%%%%%%%%%%%%%%%%%%%%%%%%%%%%%%%%%%%%%%%%%%%%%%%%%
%\newpage

\begin{abstract}
Fluidic transport in inverted T-shaped cavities with the flow entering through the top and exiting from the two bottom outlets experiences an interesting phenomenon that causes particles having density lower than that of the fluid medium to get trapped at the junction, in a horizontal formation. However, this only occurs across a small range of Reynolds numbers and that too in the laminar regime. The unexpected phenomenon is conjectured to be modulated by formation of vortex tubes in the flow. 
Interestingly enough, such T-shaped (or, more generically, Y-shaped) cavities are also seen quite widely in anatomic pathways, e.g., in animal upper airways. It can be hypothesized that this trapping phenomenon can emerge in such geometries as well.  Our current model simulates the occurrence with water as the ambient fluid medium passing through an idealized T-shaped space and measures the length of the vortices in each of the arms of the junction. In our study, we have conducted the investigation using a Reynolds number of 400 that lies in the laminar regime and have estimated the vortex tube lengths using the concept of nodal maximum velocities that the streamwise flow would attain on vortex dissipation. Vorticity and helicity contour variations as one goes further away from the junction have also been reported.

\flushright\noindent(211 words)
\end{abstract}

\vspace{-0.5cm}
\newpage

\section{Introduction}

% Figure 1
\begin{figure}[b!]
%\vspace{-0.15cm}
\begin{center}
\includegraphics[width=\textwidth]{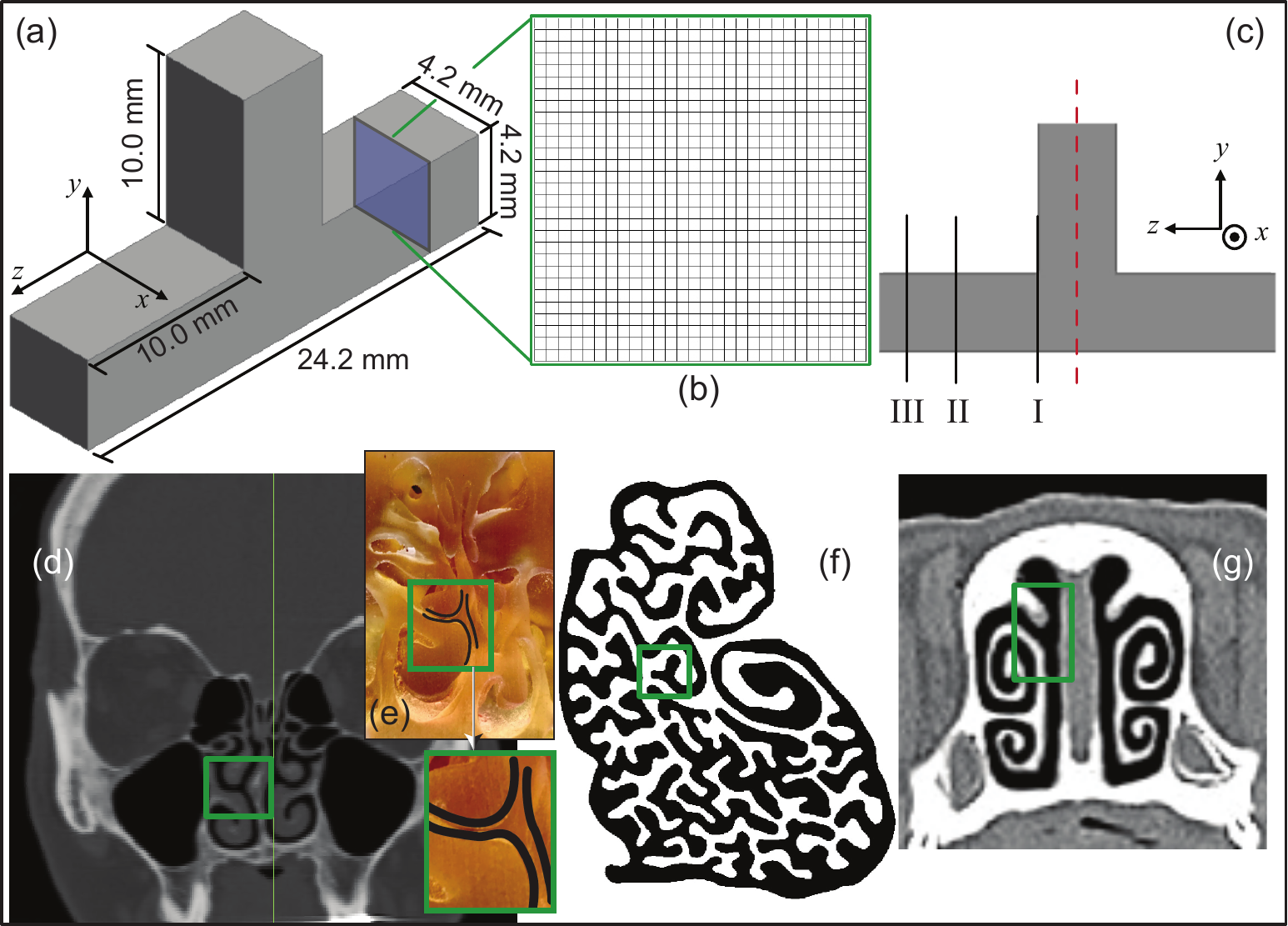}
%\vspace{-0.75cm}
\caption{
(a) T-junction geometry used in our model.
(b) Mesh profile.
(c) Defined planes for parameter calculations with the central reference plane at the junction marked in red.
(d) T and Y-shaped junctions seen in (d, e) human, (f) dog, and  (g) pig upper airway cross-sections\cite{yuk2022rsi,diab2021aps,yuk2021aps,chakraborty2020aps,yuk2020aps,chung2020aps}. The visuals are from medical-grade computed tomography (CT) scans. Panel (e) is a 3D-printed replica of a CT-based reconstruction of human upper airway. T/Y-shaped regions are subtended therein between the turbinates that line the intranasal pathway and protrude into the main airspace. The animal airway scans are courtesy of Dr.~Sunghwan Jung (Department of Biological and Environmental Engineering, Cornell University).
}
\label{fig1}
\end{center}
%\vspace{-0.5cm}
\end{figure}

A common fluid flow geometry widely observable both in nature as well as in industrial applications is a T-junction (e.g., see Figure~\ref{fig1}(a)-(c)) -- or in a somewhat more generic setting, a junction that approximately bears a Y-shape. Despite the apparent geometric simplicity of the domain, past research\cite{vigolo_unexpected_2014,chen_vortex_2017} has demonstrated how intriguing a flow physical system therein can get, with the formation of vortex tubes along the horizontal passage leading to trapping of particles that have lower density than that of the ambient fluidic medium. Remarkably enough, similar micro-channels are also copious in biological systems\cite{yuk2022rsi,diab2021aps,yuk2021aps,chakraborty2020aps,yuk2020aps,chung2020aps}, as for example in  human and animal upper respiratory pathways; see Figure~\ref{fig1}(d)-(g) for representative airway cross-sections. Vortex-induced trapping \cite{vigolo_unexpected_2014,bec_eects_nodate}, although necessary for heat and mass transfer operations especially at the anatomic length scales, has critical ramifications in such bifurcating flows because it is one of the main reasons behind particles getting trapped / slowed down, often contrary to the generally expected trends for particle trajectories in such spaces. Flows with varied particle sizes when passed through T-junctions experience a phenomenon called \textit{preferential concentration}\cite{bec_eects_nodate}, which essentially constitutes a preferential discharge of light and heavy particles in and out of vortices. As a result, the heavier particles with their higher inertia are ejected out of the vortices much sooner than the lighter ones. This gives rise to a concentrated zone of smaller particles immediately after the junction\cite{manoorkar_particle_2021}.

Complex hydrodynamic behavior is seen in junctions, often triggered by secondary Dean instabilities when Reynolds number (\textit{Re}) controlled inertial effects come into play \cite{balan_investigations_2012,zhang_investigations_2015,nivedita_dean_2017,damian_flow_2017}. The curl in velocity streamlines\cite{stigler_fluid_2012} right after the bifurcation is dependent on the inlet velocity magnitude which also governs \textit{Re}. For bifurcations with rectangular cross-sections, axisymmetric vortex breakdown is observed when it crosses a threshold \textit{Re} of around 320\cite{chan_microscopic_2018}. Vortex breakdown is basically the formation of recirculating regions after the bifurcation inside which bubbles get trapped and move in a swirling motion.\cite{mora_numerical_2019} From our simulation, it has been confirmed that the streamwise size of such vortex formations steadily decays with distance from the bifurcation and ultimately \textit{dies down}. Past research\cite{vigolo_unexpected_2014,chen_vortex_2017} has shown that particles that have specific gravity less than 1, when released in a water stream flowing inside the T-junction -- would first travel along with the mean flow till the \textit{vortex breakdown points}, i.e., the internal stagnation points inside the fluid where the particles get pushed inward back towards the bifurcation causing a `permanent' trapping scenario\cite{stone2022}. However, this occurs for a specific range of \textit{Re} values which lie well in the laminar flow regime. The present study replicates the underlying flow field in a test T-geometry and will extend the modeling approach to a biological topology, such as the ones shown in Figure \ref{fig1}(e)-(f).\\

%%%%%%%%%%%%%%%%%%%%%%%%%%%%%%%%%%%%%%%%%%%%%%%%%%%%%%%%%%%%%%%%%%%%%%%%%%%%%%%%%%%%%%%%%%%%%%%%%%%%%%%%%%%%%%%%%%%%%%%%%%%%%%%%%%%%%%%%%%%%%%%%%%%%%%%%%%%%%%%%%%%%%%%%%%%%%%%%%%%%%%%%%%%%%%%%%%%%

\section{Methods}

\subsection{Geometry building}\label{methods:geometry}
The geometry shown in Figure \ref{fig1}(b) is a representation of the test T-junction used in this study. The simulated fluidic system flows in along the negative $y$-direction, and after bifurcating, flows out along the positive and negative $z$-directions. A square cross-section with side lengths as $4.2$ mm was assumed, with each arm being $10.0$-mm long; 
%The DesignModeler module of ANSYS was used to do the same. 
see Figure \ref{fig1}(d) for a sample mesh profile with an element size of $0.15$ mm.% which was done in ICEM CFD module with physics preference set to CFD and solver preference to Fluent.

\subsection{Computational framework}\label{methods:framework}

Flow behavior was replicated with the imposed physical properties for the ambient fluid medium based on those for water at room temperature and by initializing an inlet velocity of 0.1 m/s, thereby setting the \textit{Re} for the test flow to be $\approx$ 400. This value of \textit{Re} lies comfortably below the threshold limit to show any turbulent scales\cite{selvam_flow_2016, tuoc_critical_nodate}. Our numerical scheme for the viscous-laminar flow state employed pressure-velocity coupling set to SIMPLEC and spatial discretization set to second-order upwind. Convergence criteria for the simulations was defined at $10^{-4}$ for the continuity residual and $10^{-5}$ for the velocity components.

Sixteen planes were defined starting from the section marked as I in Figure \ref{fig1}(c) which was $2.4$ mm away from the reference plane (marked in red) to monitor the changes in helicity, vorticity, and the mean streamwise velocity profiles with the planes placed 0.5 mm apart from each other. Planes II and III in Figure \ref{fig1}(c) show the tenth and sixteenth planes which would be $7.4$ mm and $10.4$ mm from the reference plane respectively.\\

%%%%%%%%%%%%%%%%%%%%%%%%%%%%%%%%%%%%%%%%%%%%%%%%%%%%%%%%%%%%%%%%%%%%%%%%%%%%%%%%%%%%%%%%%%%%%%%%

%%%%%%%%%%%%%%%%%%%%%%%%%%%%%%%%%%%%%%%%%%%%%%%%%%%%%%%%%%%%%%%%%%%%%%%%%%%%%%%%%%%%%%%%%%%%%%%%%%%%%%%%%%%%%%%%%%%%%%%%%%%%%%%%%%%%%%%%%%%%%%%%%%%%%%%%%%%%%%%%%%%%%%%%%%%%%%%%%%%%%%%%%%%%%%%%%%%%

\section{Results}
% Figure 2
\begin{figure}
%\vspace{-0.15cm}
\begin{center}
\includegraphics[width=\textwidth]{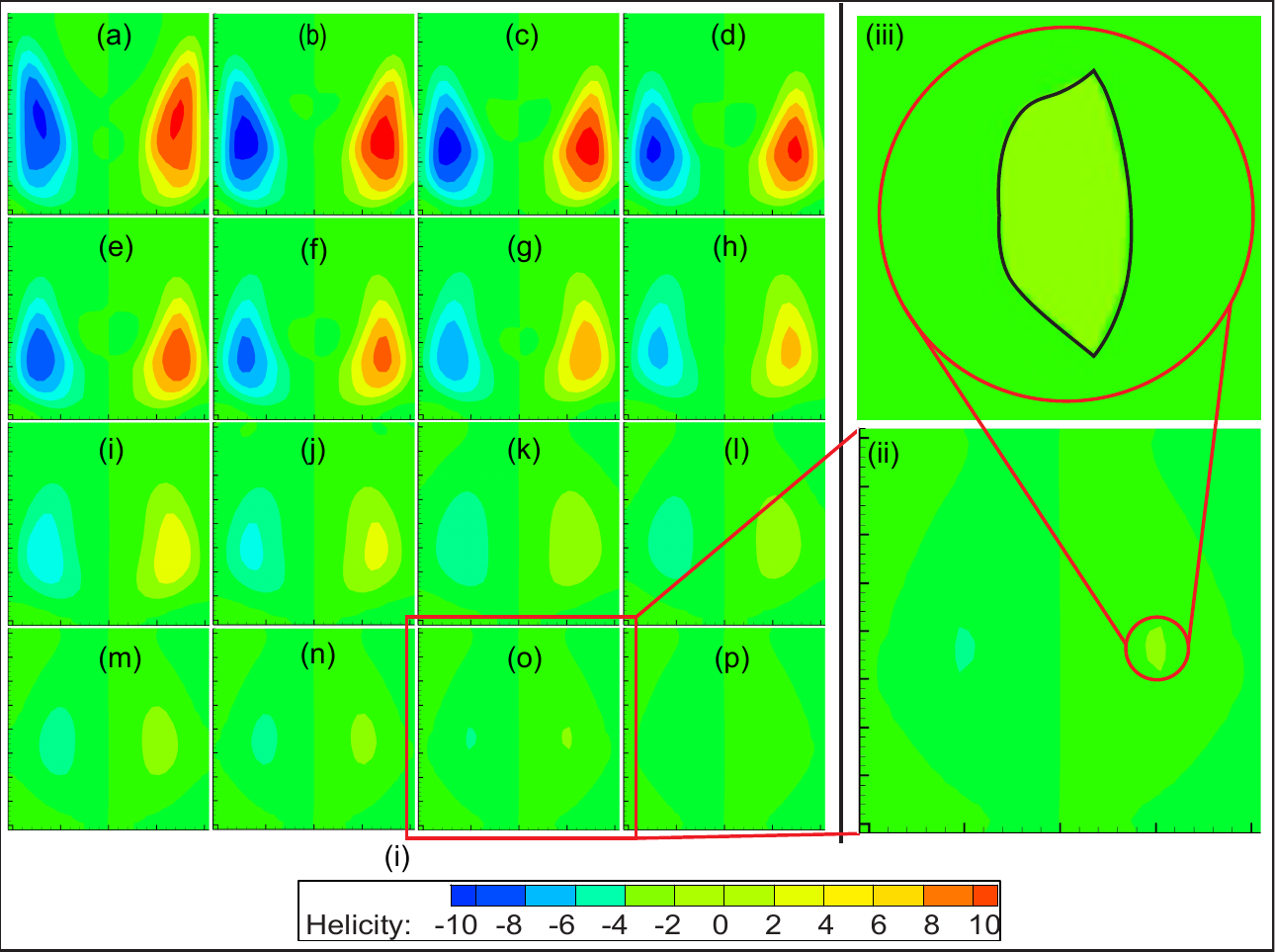}
%\vspace{-0.5cm}
\caption{(i) Helicity contour plots of sixteen planes. (ii) Magnified view of the fifteenth plane. (iii) Magnified spatial structure of helicity, as seen in (ii).}
\label{fig2}
\end{center}
%\vspace{-0.5cm}
\end{figure}
\subsection{Spatial evolution of helicity along the horizontal pathway in the streamwise direction}\label{results:helicity}
Helicity of a particle is defined by the projection of its spin on to the direction of its momentum\cite{cao_helical_2022,moffatt2014helicity}. It measures the extent to which vortex lines curl around the central axis known as the core. Figure \ref{fig2}(i) shows the helicity contour plots of the sixteen test planes in ordered alphabetical sequence. It is seen that as one moves away from the reference plane, the helicity contours decrease both in size and magnitude. This helps us quantitatively assess the vortex structures as well, as they also, in turn, decay and dissipate within the outlet pipe. The contours eventually fade out in the fifteenth plane. The vorticity plots, presented next, should also show a similar trend where the contours fade out sooner and not later.

\subsection{Spatial evolution of vorticity  along the horizontal pathway in the streamwise direction and its relationship with helicity}\label{results:vorticity}
% Figure 3
\begin{figure}
%\vspace{-0.15cm}
\begin{center}
\includegraphics[width=\textwidth]{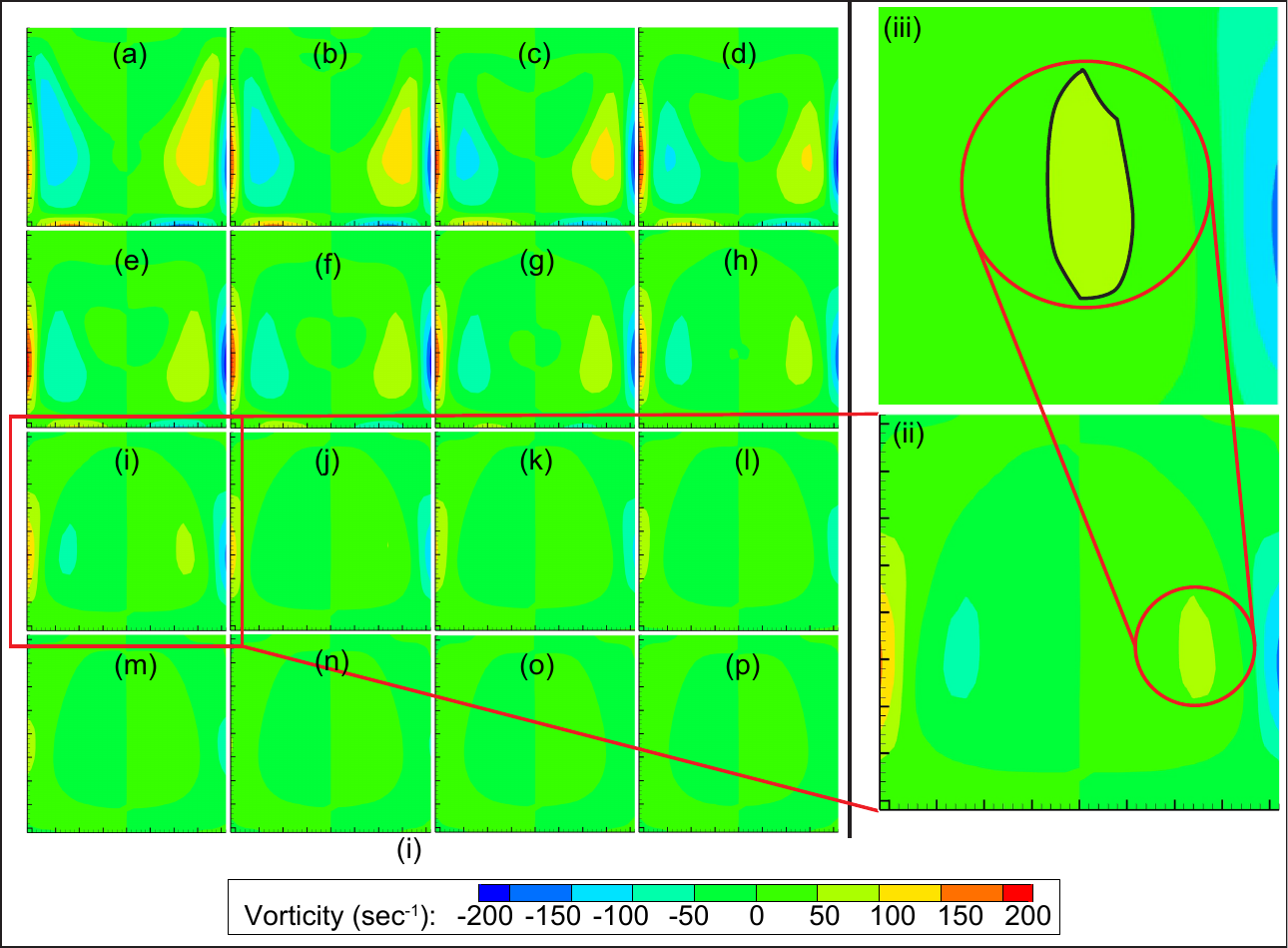}
%\vspace{-0.5cm}
\caption{(i) Vorticity contour plots of sixteen planes. (ii) Magnified view of the ninth plane. (iii) Magnified vorticity structure as seen in (ii).}
\label{fig3}
\end{center}
%\vspace{-0.5cm}
\end{figure}
Vorticity is a property of fluid flow, computed as the curl of its velocity field\cite{cao_helical_2022,cao_structure_2022,stremler_mathematical_2011-1,stremler_mathematical_2011,ellingsen_designing_2021,eyink_turbulent_2008,hufnagel_three-dimensional_2018}.
Figure~\ref{fig3} shows the vorticity contour plots. It is seen that these plots show the same decreasing trend in magnitude and size, as we saw in Section~\ref{results:helicity}. However, the vorticity completely dissipates at the ninth plane itself, which would be $6.9$ mm away from the reference plane, implying that the vortices terminate well before the helices do. Panel (ii) of Figure~\ref{fig3} is a magnified version of the ninth plane with panel (iii) showing the vortex structure before it eventually decays to have a negligible impact on the surrounding flow field.

\subsection{Spatial analysis of the mean streamwise velocity trend to locate the vortex termination region}\label{results:velocity}
\begin{figure}[t]
%\vspace{-0.15cm}
\begin{center}
\includegraphics[width=\textwidth]{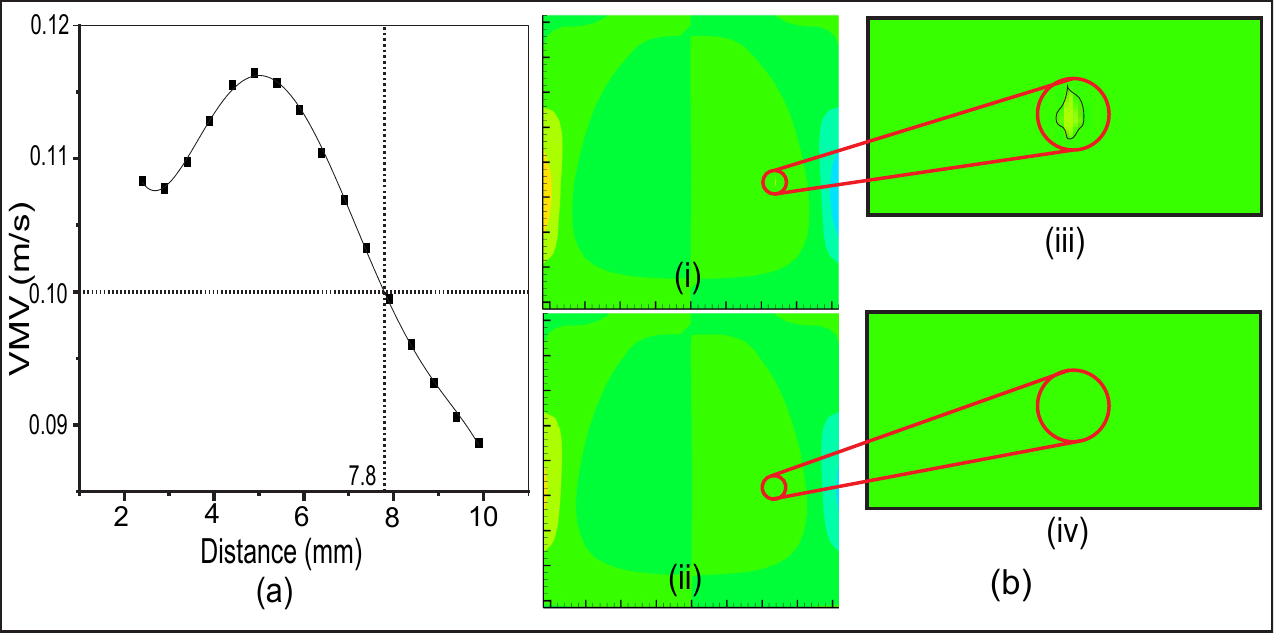}
%\vspace{-0.5cm}
\caption{(a,b) Vertex Maximum Velocity (VMV) versus distance from the central plane, marked red in Figure~\ref{fig1}(c). (b) Zoomed-in vorticity contour plots mapped across the tenth and eleventh planes.}
\label{fig4}
\end{center}
%\vspace{-0.5cm}
\end{figure}

Magnitudes of vertex maximum velocities (VMV) were noted for all the planes and it was seen in Figure~\ref{fig4}(a) that they increased to a peak velocity and then started decreasing monotonically. A recall to Section~\ref{methods:framework} would show that the inlet velocity was taken to be 0.1 m/s. Therefore, it can be postulated that the next instance where the VMV reaches and then ultimately falls below 0.1 m/s again would be the approximate position where the vortex \textit{dies down}. Hence, a plot between VMV and distance from reference plane was made from which we were able to figure out graphically that this location is $7.8$ mm from the red-marked reference plane, which would be somewhere in between the eleventh and twelfth plane. Refer to Table~\ref{tab1} for the related data.

%Velocity profile table
% \usepackage{multirow}
% \usepackage{graphicx}
\begin{table}
\centering
\resizebox{\textwidth}{!}{%
\renewcommand{\arraystretch}{0.85}
\begin{tabular}{|c|c|c|}
\hline
                        &                                     &                                            \\
\multirow{-2}{*}{\textbf{Plane}} & \multirow{-2}{*}{\textbf{Distance {(}mm{)}}} & \multirow{-2}{*}{\textbf{Vertex Maximum Velocity {(}m/s{)}}} \\ \hline
1                       & 2.4                                 & 0.1083                                 \\ \hline
2                       & 2.9                                 & 0.1077                                 \\ \hline
3                       & 3.4                                 & 0.1097                                  \\ \hline
4                       & 3.9                                 & 0.1128                                 \\ \hline
5                       & 4.4                                 & 0.1155                                 \\ \hline
6                       & 4.9                                 & 0.1164                                 \\ \hline
7                       & 5.4                                 & 0.1156                                  \\ \hline
8                       & 5.9                                 & 0.1136                                 \\ \hline
9                       & 6.4                                 & 0.1104                                 \\ \hline
10                      & 6.9                                 & 0.1068                                 \\ \hline
\rowcolor[HTML]{D3D3D3} 
11                      & 7.4                                 & 0.1033                                 \\ \hline
\rowcolor[HTML]{D3D3D3} 
12                      & 7.9                                 & 0.0994                                \\ \hline
13                      & 8.4                                 & 0.0960                                \\ \hline
14                      & 8.9                                 & 0.0931                                \\ \hline
15                      & 9.4                                 & 0.0906                                \\ \hline
16                      & 9.9                                 & 0.0886                                \\ \hline
\end{tabular}%
}

\caption{The first column lists the plane numbers, with the subsequent columns listing the data on the distance from reference plane and the VMV, for each of the planar cross-sections through the flow field.}
\label{tab1}
\vspace{1cm}
\end{table}

% Figure 4

\begin{wrapfigure}{r}{7.8cm}
\vspace{-0.9cm}
\begin{center}
\includegraphics[width=7.8cm]{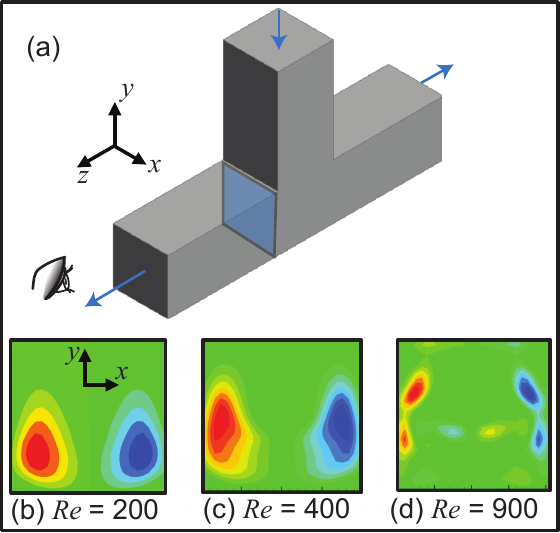}
%\vspace{-0.75cm}
\caption{(a) Test domain with blue arrows indicating the mean direction of the simulated water flow. Panel (b)-(d) show the helicity contours mapped across the blue cross-section shown in (a). The flow structures emerge and evolve beyond a critical \textit{Re}, with an eventual local decay as the \textit{Re} is raised. The results agree with prior findings\cite{vigolo_unexpected_2014,chen_vortex_2017}.}\label{fig5}
\end{center}
\vspace{-0.75cm}
\end{wrapfigure}

Now, if we go back to Figure~\ref{fig3} and extract the tenth and eleventh (Panel (j) and (k) respectively) planes and zoom-in as shown in Figure~\ref{fig4}(b) we can see that there still exists a minute vortex structure in Panel (i), i.e.~at the tenth plane, which disappears in the very next plane, i.e.~in Panel (ii). So, it is safe to conclude that the vortex does eventually \textit{die down}  in the near vicinity where the VMV equals the inlet velocity. This is, therefore, the point where \textit{vortex breakdown} occurs and the particles might be forced back inwards into the upstream vortex tube,  thus trapping them until the \textit{Re} is raised further which leads to a local dissipation of the coherent vortex structures (see e.g., Figure~\ref{fig5}).\\

%%%%%%%%%%%%%%%%%%%%%%%%%%%%%%%%%%%%%%%%%%%%%%%%%%%%%%%%%%%%%%%%%%%%%%%%%%%%%%%%%%%%%%%%%%%%%%%%%%%%%%%%%%%%%%%%%%%%%%%%%%%%%%%%%%%%%%%%%%%%%%%%%%%%%%%%%%%%%%%%%%%%%%%%%%%%%%%%%%%%%%%%%%%%%%%%%%%%

\section{Discussion}

%\begin{itemize}[leftmargin=*]

Investigations on vortex-induced trapping in biomedical pathways, modeled after the explored system, could yield significant inputs for intranasal drug delivery\cite{basu2020scirep,basu2018ijnmbe} and breathing problems\cite{thomas2013particle}. Bifurcating flows are very common in biological scenarios. Animal anterior airways\cite{basu_computational_2021,yuk2022rsi} have multiple locations where such T or Y-shaped junctions are prominent (Figure~\ref{fig1}(d)). Fluid friction within these channels are responsible for heat generation along the inner walls\cite{lintermann_investigations_2011}. Hence, a study of particle trapping within the vortex tubes inside such topologies is evidently an important aspect to explore for such bifurcating flows\cite{inthavong_computational_2009,pan_marangoni_2012,li_nasal_2018}. To advance the state-of-art, we will shortly supplement this study with a model for the flow behavior in T/Y-shaped animal respiratory morphology reconstructed medical imaging data.

From our evaluation, it is seen that each of the vortices is ranging over a length that occupies approximately 50-55\% of the total outlet arm for the set of boundary conditions used. Additionally, vortex breakdown point is correlated to the inlet velocity; the more the inlet velocity, the farther the vortex breakdown point will be from the junction.

Particle tracking is an important aspect which will help us figure out how long any particle in the fluid stream might get entrapped within the vortex bubble. Categorizing their properties such as density, size, shape would also become necessary then. Dependency of the angle between junctions (e.g., if we consider Y-junctions) can play a vital role in how the vortex structures and their spatial extent evolve\cite{noori_calculation_2022}. Multiphase fluidic behavior as well as complex rheologies can also affect the particle trapping property of the vortex bubbles\cite{kim_inertio-elastic_2017,park_vortex_2019,xue_vortex_2022}. For example, the symmetric occurrence of the vortex bubbles that we observed in our case does not stand in visco-elastic fluids\cite{poole_symmetry-breaking_2014}. In the context of biomedical transport, note that a typical arterial flow, based on the time scale in consideration, is pulsatile in nature, but we modeled the test system as a steady flow throughout the entire duration of our simulation. Hence, to realistically model an arterial flow, one has to make appropriate time-dependent changes to the inlet flow conditions\cite{cox_effect_2021}. A power law-based correlation between \textit{Re} and vorticity has been implemented through use of Random Vortex Method (RVM)~\cite{noori_use_2022}; the same can be done with our model to compare between the two parameters.\\

%\end{itemize}

%\section{Final takeaways}

%%%%%%%%%%%%%%%%%%%%%%%%%%%%%%%%%%%%%%%%%%%%%%%%%%%%%%%%%%%%%%%%%%%%%%%%%%%%%%%%%%%%%%%%%%%%%%%%%%%%%%%%%%%%%%%%%%%%%%%%%%%%%%%%%%%%%%%%%%%%%%%%%%%%%%%%%%%%%%%%%%%%%%%%%%%
%\vspace{1cm}
\noindent\hrulefill
%\vspace{-0.25cm}

%\section*{Acknowledgements}
%The authors thank the \textit{Bioinspiration \& Biomimetics} editorial board for the invitation to publish. This material is based on work partially supported by the National Science Foundation (NSF) RAPID Grant 2028069. Any opinions, findings, and conclusions or recommendations expressed here are, however, those of the authors and do not necessarily reflect views of the NSF. 

\section*{Author contributions}
SB:~conceptualization, project administration, writing.
SD:~digital reconstructions, simulations, theoretical calculations, data analysis, writing.

\section*{Data availability}
The simulation data-sets and scripts for the plotted figures -- are available on-request from the corresponding author, through a shared-access Google Drive folder.

\section*{ORCID iDs}
Saikat Basu~~\href{https://orcid.org/0000-0003-1464-8425}{https://orcid.org/0000-0003-1464-8425}\\
Sneham Das~~\href{https://orcid.org/0000-0002-4722-6272}{https://orcid.org/0000-0002-4722-6272}\\

%\vspace{0.5cm}
\noindent\hrulefill\\
%\newpage

%%%%%%%%%%%%%%%%%%%%%%%%%%%%%%%%%%%%%%%%%%%%%%%%%%%%%%%%%%%%%%%%%%%%%%%%%%%%%%%%%%%%%%%%%%%%%%%%%%%%%%%%%%%%%%%%%%%%%%%%%%%%%%%%%%%%%%%%%%%%%%%%%%%%%%%%%%%%
\vspace{0.5cm}
%\newpage 
\noindent\textbf{References}\\

\bibliographystyle{unsrt}
\bibliography{Ref}

\end{document}